\documentclass[%
 reprint,
superscriptaddress,
 amsmath,amssymb,
 aps,
]{revtex4-1}

\usepackage{graphicx}% Include figure files
\usepackage{dcolumn}% Align table columns on decimal point
\usepackage{bm}% bold math
\begin{document}

\preprint{APS/123-QED}

\title{The $\gamma$-ray Strength Function for Thallium Isotopes relevant to \\the $^{205}$Pb--$^{205}$Tl Chronometry}% Force line breaks with \\

\author{H. Utsunomiya}
\affiliation{Konan University, Department of Physics, 8-9-1 Okamoto, Higashinada, Japan}
\email{hiro@konan-u.ac.jp}

\author{T.~Renstr{\o}m}
\affiliation{Department of Physics, University of Oslo, N-0316 Oslo, Norway}

\author{G.~M.~Tveten}
\affiliation{Department of Physics, University of Oslo, N-0316 Oslo, Norway}

\author{S.~Goriely}
\affiliation{Institut d'Astronomie et d'Astrophysique, Universit\'{e} Libre de Bruxelles, Campus de la Plaine, CP-226, 1050 Brussels, Belgium}

\author{T.~Ari-izumi}
\affiliation{Konan University, Department of Physics, 8-9-1 Okamoto, Higashinada, Japan}

\author{D.~Filipescu}
\affiliation{National Institute for Physics and Nuclear Engineering, Horia Hulubei (IFIN-HH), 407 Atomistilor Str., P.O. Box MG6, Bucharest-Magurele, Romania}

\author{J.~Kaur}
\affiliation{Extreme Light Infrastructure Nuclear Physics, "Horia Hulubei" National Institute for Physics and Nuclear Engineering (IFIN-HH), 30 Reactorului, 077125 Bucharest-Magurele, Romania}

\author{Y.-W.~Lui}
\affiliation{Cyclotron Institute, Texas A\& M University, College Station, Texas 77843, USA}

\author{W.~Luo}
\affiliation{School of Nuclear Science and Technology, University of South China, Hengyang 421001, China}

\author{S.~Miyamoto}
\affiliation{Laboratory of Advanced Science and Technology for Industry, University of Hyogo, 3-1-2 Kouto, Kamigori, Ako-gun, Hyogo 678-1205, Japan}

\author{A.~C.~Larsen}
\affiliation{Department of Physics, University of Oslo, N-0316 Oslo, Norway}

\author{S.~Hilaire}
\affiliation{CEA, DAM, DIF, F-91297 Arpajon, France}

\author{S.~P\'{e}ru}
\affiliation{CEA, DAM, DIF, F-91297 Arpajon, France}

\author{A.~J.~Koning}
\affiliation{Nuclear Data Section, International Atomic Energy Agency, A-1400 Vienna, Austria}

\date{\today}% It is always \today, today,

\begin{abstract}
Photoneutron cross sections were measured for $^{203}$Tl and $^{205}$Tl at energies between the
one- and two-neutron thresholds using quasi-monochromatic $\gamma$-ray beams produced in laser
Compton-scattering at the NewSUBARU synchrotron radiation facility.
Our new measurement results in cross sections significantly different from the previously reported bremsstrahlung experiment, leading to rather different GDR parameters, in particular to lower GDR peak energies and higher peak cross sections. 
The photoneutron data are used to constrain the $\gamma$-ray strength function on the basis of the Hartree-Fock-Bogolyubov plus quasi-particle
random phase approximation using the Gogny D1M interaction.
Supplementing the experimentally constrained $\gamma$-ray strength function with the zero-limit E1 and M1 contributions for the de-excitation mode, we estimate the Maxwellian-averaged cross section for the s-process branching-point  nucleus $^{204}$Tl in the context of the $^{205}$Pb -- $^{205}$Tl chronometry.

\end{abstract}

%\pacs{Valid PACS appear here}
\maketitle
\section{Introduction}
In recent years, one witnesses a rapid growth of experimental and theoretical studies of the $\gamma$-ray strength function ($\gamma$SF) \cite{Bartholomew73,Lone85,RIPL3} across the chart of nuclei. The $\gamma$SF in the de-excitation mode, which is the nuclear statistical quantity equivalent to the transmission coefficient of the $\gamma$-ray emission, is a key quantity to determine radiative neutron capture cross sections that are of direct relevance to the nucleosynthesis of elements heavier than iron. It  may be a highlight of the recent development \cite{Goriely18a} to have reached a recognition that goes beyond the Brink hypothesis \cite{Brink,Axel}; the $\gamma$SF in de-excitation mode differs from that in excitation mode in the zero-limit behavior of both E1 and M1 strengths, the latter of which referred to as upbend was experimentally observed \cite{Voin04,Gutt05,Algi08} and theoretically supported by the shell-model calculation \cite{Schw13,Brow14,Siej17a,Siej17b,Kara17,Schw17}.    

\begin{figure}
\includegraphics[bb = 150 100 550 425, scale=0.4]{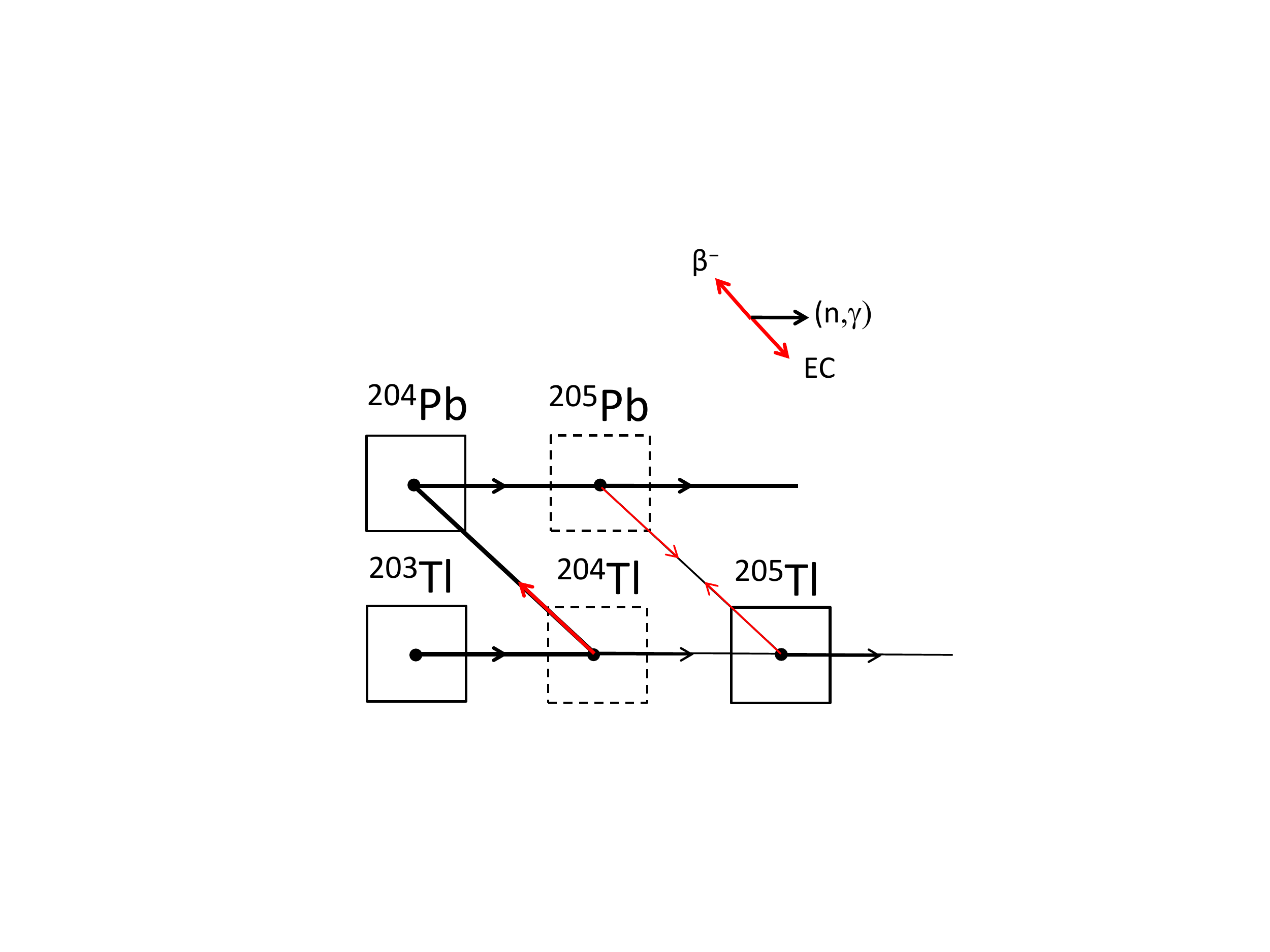}
\caption{(Color online)  An excerpt of the chart of nuclei depicting the Tl-Pb region in the s-process path.}
\label{fig:chart}
\end{figure}

The present research interest in the $\gamma$SF lies in the thallium isotopes in relation to the s-process nucleosynthesis.    
The s-process in the Tl-Pb region involves a possible astrophysical application called the $^{205}$Pb--$^{205}$Tl chronometry as depicted in Fig.~\ref{fig:chart}.  
This chronometer relies on the production and survival of a short-lived $^{205}$Pb in certain s-process conditions like low-mass Asymptotic Giant Branch stars \cite{Mowl89} and massive Wolf-Rayet stars \cite{Arno97} and possible isotopic anomalies in meteoritic Tl due to in-situ decay of now extinct $^{205}$Pb \cite{Ande60,Osti69,Huey72}. The $^{205}$Pb is produced by the s-process only via $^{204}$Tl, an s-process branching-point nucleus with a half-life of 3.78 yr.     
While the $^{205}$Pb decays to $^{205}$Tl via electron capture with a half-life of 1.7 $\times$ 10$^7$ yr in laboratory conditions, the electron capture is accelerated in s-process conditions by the thermal populations of low-lying nuclear excited states \cite{Taka83}.  Moreover, the $^{205}$Tl, when highly ionized, undergoes the so-called bound-state $\beta^{-}$-decay to $^{205}$Pb \cite{Daud47,Bahc61,Taka83,Yoko85}. With the nuclear physics and astrophysics behind,  the $^{205}$Pb--$^{205}$Tl chronometer may determine the time span between the last nucleosynthetic events that modified the composition of the solar nebula and the formation of the solar system solid bodies \cite{Yoko85}.    

The s-process production of $^{205}$Tl and $^{205}$Pb depends on the $^{204}$Tl radiative neutron capture cross section (Fig.~\ref{fig:chart}).
In the absence of possible direct measurements, we present an experimentally constrained estimate of the  $^{204}$Tl(n,$\gamma$) cross section obtained with the $\gamma$-ray strength function method \cite{Utsu10a,Utsu10b,Utsu11,Utsu13,Fili14,Nyhu15}. The $\gamma$SF from the Hartree-Fock-Bogolyubov plus quasi-particle random phase approximation (QRPA) based on the Gogny D1M interaction \cite{Martini16,Goriely16b,Goriely18a} for both E1 and M1 components is constrained to new experimental Tl photoneutron cross sections. In Sect.~\ref{sec_exp}, our experimental procedure is described and in Sect.~\ref{sec_data} data are analysed. Our resulting photoneutron cross sections are discussed in Sect.~\ref{sec_disc} and compared with D1M+QRPA calculations. Based on the same nuclear ingredients, the calculated radiative neutron capture of the stable  $^{203}$Tl and  $^{205}$Tl are compared with experimental data, before applying the same procedure to the estimate of the $^{204}$Tl radiative neutron capture cross section. Finally conclusions are drawn in Sect.~\ref{sec_conc}.

\section{Experimental procedure} 
\label{sec_exp}
\begin{figure*}
%\center

\includegraphics[bb = 300 100 900 400, scale=0.3]{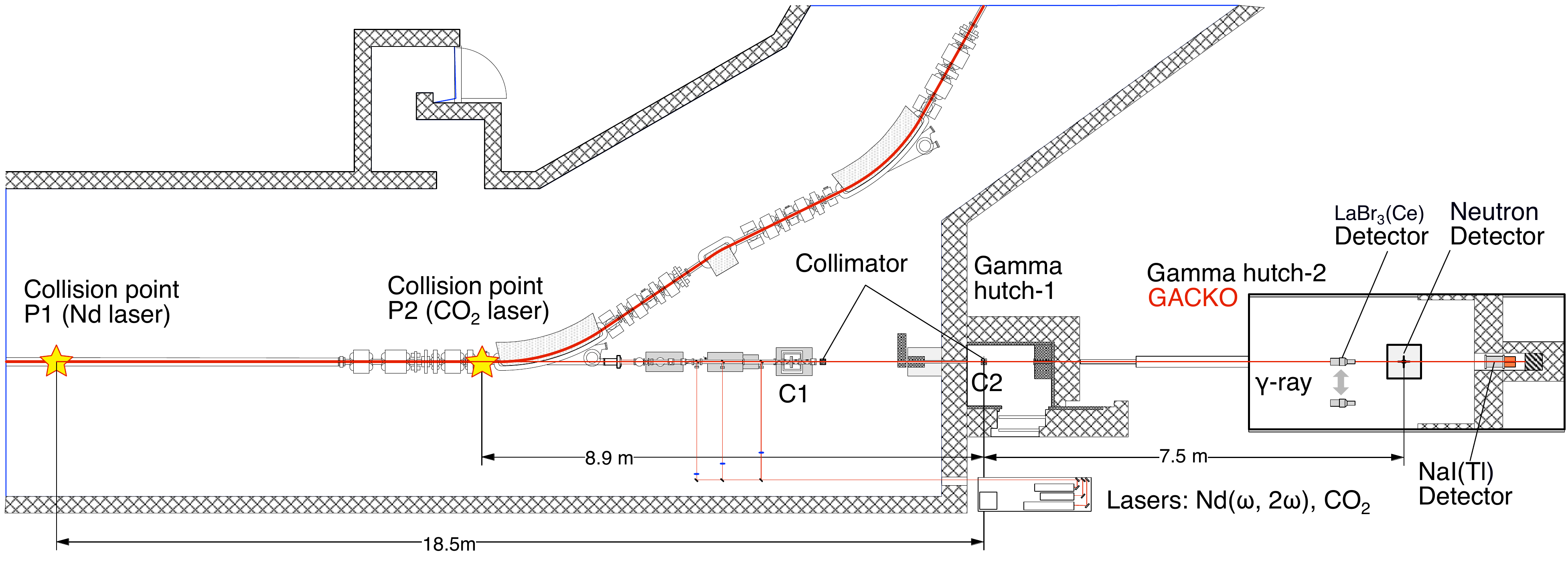}
\vspace{1.2cm}
\caption{(Color online) A schematic illustration of the experimental set up at NewSUBARU used in the ($\gamma$, n) cross-section measurements.}
\label{fig:setup}
\end{figure*}

The photo-neutron measurements on $^{203,205}$Tl took place at the NewSUBARU synchrotronic radiation facility. Figure~\ref{fig:setup} shows a schematic illustration of the gamma-ray beam line and experimental set up. Quasi-monochromatic $\gamma$-ray beams were produced through laser
Compton scattering (LCS) of 1064 nm photons in head-on collisions with relativistic electrons. Throughout the experiment, the laser was periodically on for 80 ms and off for 20 ms, in order to measure background neutrons and $\gamma$-rays. The electrons were injected from a linear accelerator into the NewSUBARU storage ring with an initial energy of 974 MeV, then subsequently decelerated to nominal energies in the region from 651 and 664~MeV to
882 and 904~MeV, providing LCS $\gamma$-ray beams corresponding to $S_n$ up to $S_{2n}$ for $^{205}$Tl and $^{203}$Tl respectively. 
In total, 15 individual $\gamma$ beams were produced for both $^{203}$Tl and $^{205}$Tl. 
The electron beam energy has been calibrated with the accuracy on the order of 10$^{-5}$ \cite{Utsu14}.  
The energy is reproduced in every injection of an electron beam from a linear accelerator to
the storage ring. The reproducibility of the electron energy is assured in the deceleration
down to 0.5 GeV and acceleration up to 1.5 GeV by an automated control of the electron beam-optics parameters.

The energy profiles of the produced $\gamma$-ray beams were measured with a $3.5^{\prime\prime}\times 4.0^{\prime\prime}$ LaBr$_3$:Ce (LaBr$_3$) detector. The measured LaBr$_3$ spectra were reproduced by a GEANT4 code~\cite{Ioana_thesis, geant4ref} that incorporated the kinematics of the LCS process, including the beam emittance and the interactions between the LCS beam and the LaBr$_3$ detector. In this way we were able to simulate the incoming energy profile of the $\gamma$ beams with the maximum energies accurately determined by the calibrated electron beam energy. 

The $^{203,205}$Tl targets were in metallic form with an areal density of $2.693$~g/cm$^2$ and $3.978$~g/cm$^2$, respectively.
The corresponding enrichment of the two isotopes were $97.2\%$ and $99.9\%$. The target material was pressed, thanks to this metal being malleable, into uniform disks and placed inside open cylinders of aluminum. For neutron detection, the high-efficiency $4\pi$ detector was used, consisting of 20 $^3\rm{He}$ proportional counters, arranged in three concentric rings and embedded in a 36 $\times$ 36 $\times$ 50 cm$^3$ polyethylene neutron moderator~\cite{neutrondet}. The ring ratio technique, originally developed by Berman~\cite{Berman_ring_ratio}, was used to determine the average energy of the neutrons from the ($\gamma$,n) reactions. The efficiency of the neutron detector varies with the average neutron energy. The efficiency was measured with a calibrated $^{252}$Cf source and the energy dependence was determined by Monte Carlo simulations. The efficiency of the neutron detector was simulated using isotropically distributed, mono-energetic neutrons. The simulation performed for s- and p-wave neutrons shows a strong smearing effect on highly-anisotropic p-wave neutrons due to the thermalization of neutrons in the polyethylene moderator, resulting in a nearly identical efficiency for s- and p-wave neutrons \cite{Utsu2017}.
%Once the neutron detection efficiency for a given beam energy has been determined, we are able to deduce the number of ($\gamma$, n) reactions that took place during each run. 

The LCS $\gamma$-ray flux was monitored by a $8^{\prime\prime}\times 12^{\prime\prime}$ NaI:Tl (NaI)
detector during neutron measurement runs with 100$\%$ detection efficiency for the beam energies used in this experiment. The number of incoming $\gamma$ rays per measurement was estimated using the pile-up/Poisson-fitting technique described in Ref.~\cite{Kondo2011,utsunomiya2018}.

\begin{figure}[]
\includegraphics[clip,width=0.85\columnwidth]{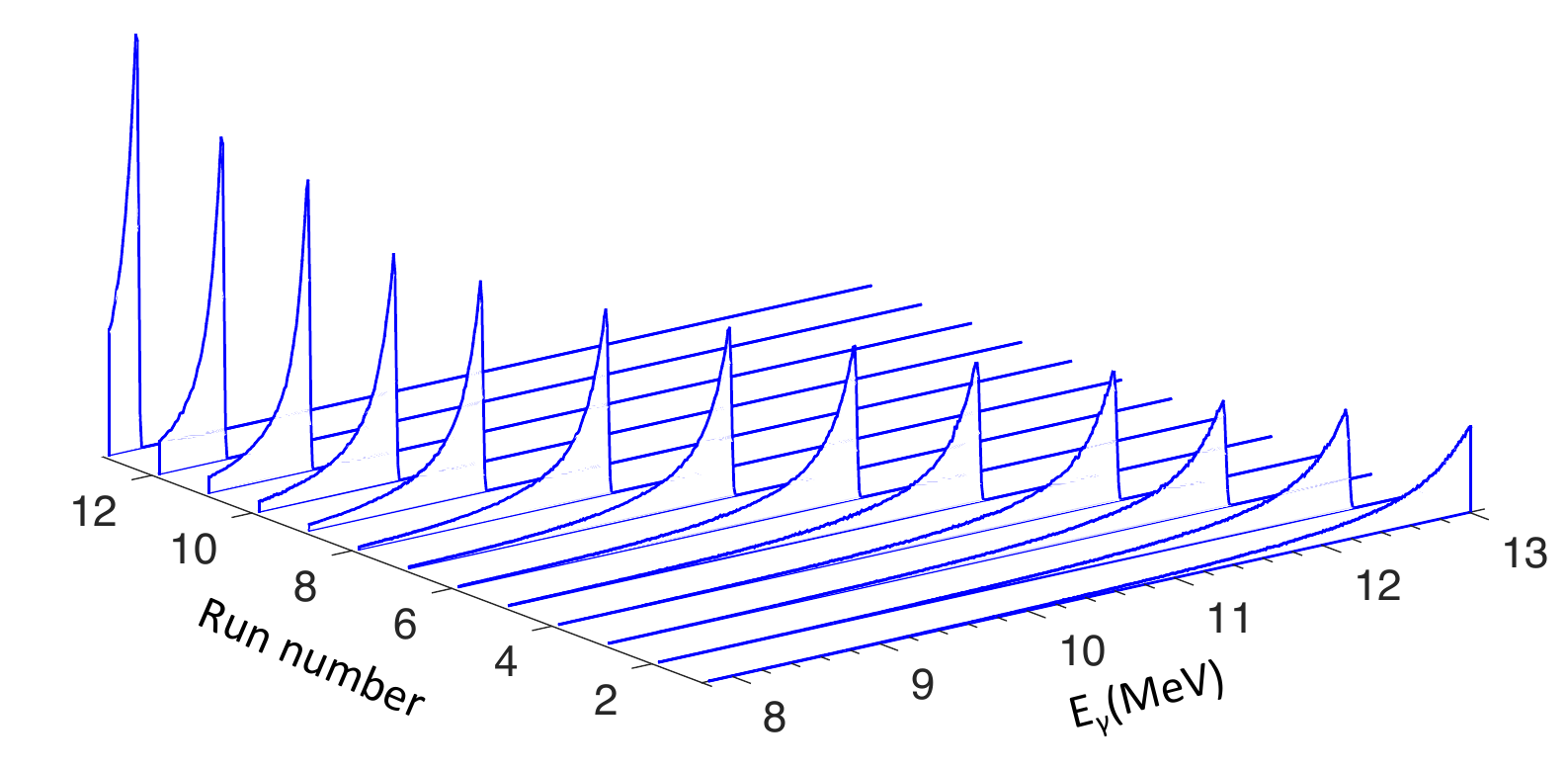}
\caption{(Color online) The simulated energy profiles for the $\gamma$-beams used in the $^{203,205}$Tl measurements.}
\label{fig:GammaProfile}
\end{figure}

The measured photo-neutron cross section for an incoming beam with maximum $\gamma$-energy $E_{\rm max}$ is given by the convoluted cross section,
\begin{equation}
\sigma^{E_{\rm max}}_{\rm exp}=\int_{S_n}^{E_{\rm max}}D^{E_{\rm max}}(E_{\gamma})\sigma(E_{\gamma})dE_{\gamma}=\frac{N_n}{N_tN_{\gamma}\xi\epsilon_n g}.
\label{eq:cross1}
\end{equation}
Here, $D^{E_{\rm max}}$ is the normalized,$\int_{S_n}^{E_{\rm max}} D^{E_{\rm max}}dE_{\gamma}= 1$, energy distribution of
the $\gamma$-ray beam obtained from GEANT4 simulations. The simulated profiles of the $\gamma$ beams, $D^{E_{\rm max}}$, used to investigate $^{205}$Tl are shown in Fig.~\ref{fig:GammaProfile}. Furthermore, $\sigma(E_{\gamma})$ is the true photo-neutron cross section as a function of energy. The quantity $N_n$ represents the number of neutrons detected, $N_t$ gives the number of target nuclei per unit area, $N_{\gamma}$ is the number of $\gamma$ rays incident on target, $\epsilon_n$ represents the neutron detection efficiency, and finally $\xi=(1-e^{\mu t})/(\mu t)$ gives a correction factor for self-attenuation in the target. The factor $g$ represents the fraction of the $\gamma$ flux above $S_n$. 

We have determined the convoluted cross sections $\sigma^{E_{\rm max}}_{\rm exp}$ given by Eq.~(\ref{eq:cross1}) for $\gamma$ beams with maximum energies in the range $S_{n}\leq E_{\rm max} \leq$ 13 MeV. The convoluted cross sections $\sigma^{E_{\rm max}}_{\rm exp}$ are not connected to a specific $E_{\gamma}$, and we choose to plot them as a function of $E_{\gamma \rm max}$. The convoluted cross sections of the two Tl-isotopes, which are often called monochromatic cross sections, are shown in Fig.~\ref{fig:MonocrossBoth}. The error bars in Fig.~\ref{fig:MonocrossBoth} represent the total uncertainty in the quantities comprising Eq.~(\ref{eq:cross1}) and consists of $\sim 3.2\%$ from the efficiency of the neutron detector, $\sim 3\%$ from the pile-up method that gives the number of $\gamma$-rays, and the statistical uncertainty in the number of detected neutrons. The statistical error ranges between $\sim$ 14 $\%$ close to neutron threshold and 0.5 $\%$ for higher $\gamma$ energies. Except for the first few data points close to separation energy, the total error is dominated by the uncertainty stemming from the pile-up method and from the simulated efficiency of the neutron detector.
For the total uncertainty, we have added these uncorrelated errors quadratically.

%---------------------------------------------------%
%---------------------------------------------------%

\section{Data analysis}
\label{sec_data}
The challenge now is to extract the deconvoluted, $E_{\gamma}$ dependent, photo-neutron cross section, $\sigma(E_{\gamma})$, from the integral of Eq.~(\ref{eq:cross1}).
Each of the measurements characterized by the beam energy, $E_{\rm max}$, correspond to folding of $\sigma(E_{\gamma})$ with the 
measured beam profile, $D^{E_{\rm max}}$.  
%---------------------------------------------------%
\begin{figure}[]
\includegraphics[clip,width=0.85\columnwidth]{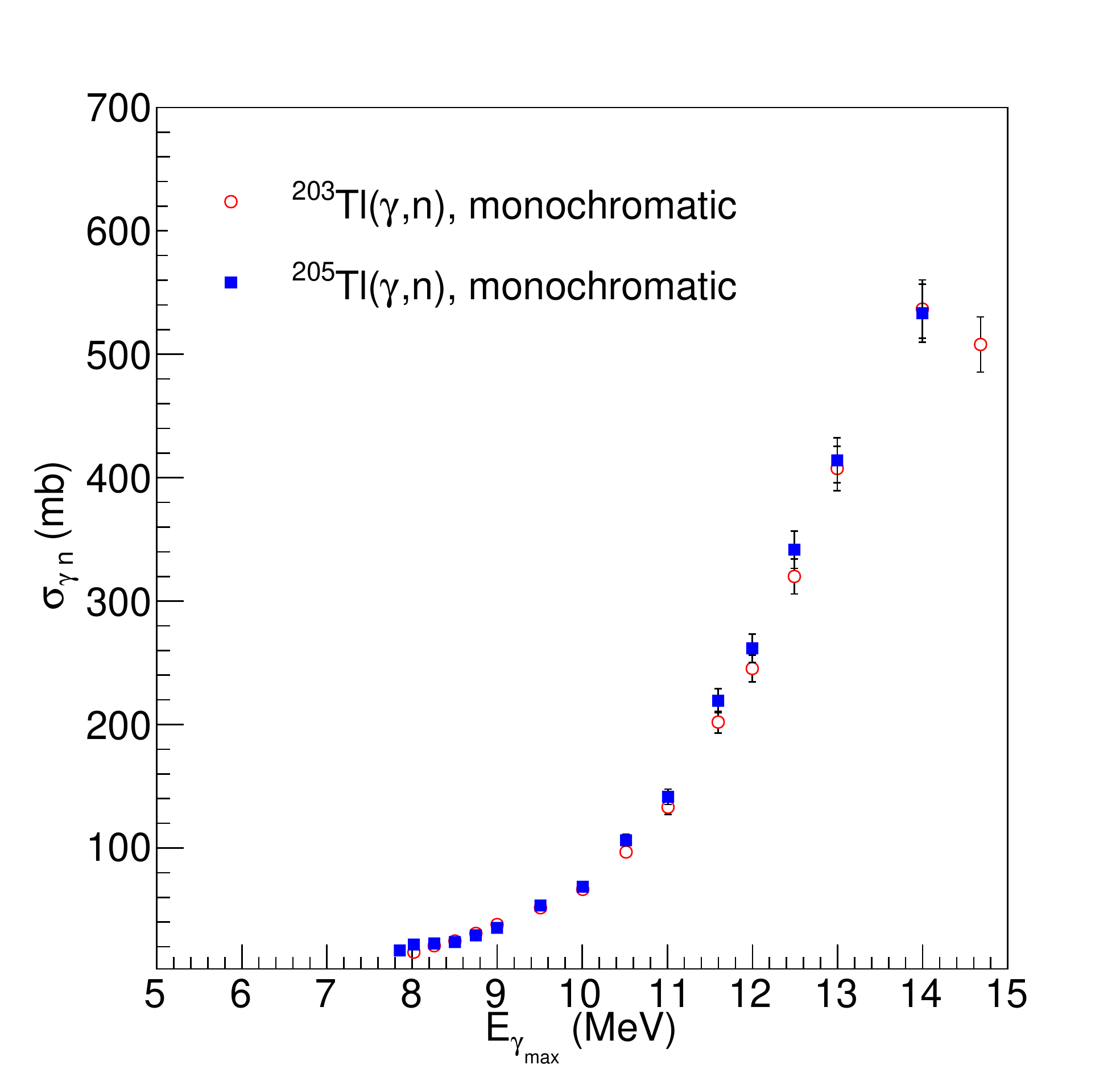}
\caption{(Color online) Monochromatic cross sections of $^{203}\rm{Tl}$ (green open circles) and $^{205}\rm{Tl}$ (blue filled squares). The error bars contain statistical uncertainties from the number of detected neutrons, the uncertainty in the efficiency of the neutron detector and the uncertainly in the pile-up method used to determine the number of incoming $\gamma$'s on target.}
\label{fig:MonocrossBoth}
\end{figure}
%---------------------------------------------------%
By approximating the integral in Eq.~(\ref{eq:cross1}) with a sum for each $\gamma$-beam profile, we are able to express the problem as a set of linear equations
\begin{equation}
\sigma_{\rm f }=\bf{D}\sigma,
\end{equation}
where $\sigma_{\rm f}$ is the cross section folded with the beam profile {\bf D}.  
The indexes $i$ and $j$ of the matrix element $D_{i,j}$ corresponds to $E_{\rm max}$ and $E_{\gamma}$, respectively.
The set of equations is given by
\begin{equation}
\begin{pmatrix}\sigma_{\rm{1}}\\\sigma_{\rm{2}}\\ \vdots \\ \sigma_N \end{pmatrix}_{\rm f}\\\mbox{}=
\begin{pmatrix}D_{ 11} & D_{ 12}& \cdots &\cdots &D_{ 1M} \\ D_{ 21} & D_{ 22}&
\cdots & \cdots &D_{ 2M} \\ \vdots &\vdots & \vdots & \vdots & \vdots \\ D_{ N1} & D_{ N2}& \cdots & \cdots &D_{ NM}\end{pmatrix}
\begin{pmatrix}\sigma_{1}\\\sigma_{2}\\ \vdots \\ \vdots \\\sigma_{M} \end{pmatrix}.
\label{eq:matrise_unfolding}
\end{equation}
Each row of $\bf{D}$ corresponds to a GEANT4 simulated $\gamma$
beam profile belonging to a specific measurement characterized by $E_{\rm max}$.  See Fig.~\ref{fig:GammaProfile} for a visual representation of the response matrix $\bf{D}$. It is clear that $\bf{D}$ is highly asymmetrical.
As mentioned, we have used $N=15$ beam energies when investigating $^{205}$Tl, but the beam profiles above $S_n$ is simulated for $M= 250$ $\gamma$ energies.
As the system of linear equations in Eq.~(\ref{eq:matrise_unfolding}) is under-determined, the true $\sigma$ vector cannot be extracted by matrix inversion. In order to find $\sigma$, we utilize a folding iteration method. The main features of this method are as follows:

\begin{itemize}

\item [1)] As a starting point, we choose for the 0th iteration, a constant trial function $\sigma^0$.
This initial vector
is multiplied with $\bf{D}$, and we get the 0th folded vector $\sigma^0_{\rm f}= {\bf D} \sigma^{0}$.
\item[2)] The next trial input function, $\sigma^1$, can be established by adding the difference of
the experimentally measured spectrum, $\sigma_{\rm{exp}}$, and the folded spectrum, $\sigma^0 _{\rm f}$,
to $\sigma^0$. In order to be able to add the folded and the input vector together, we first perform a spline
interpolation on the folded vector, then interpolate so that the two vectors have equal dimensions. Our new input vector is:

\begin{equation}
\sigma^1 = \sigma^0 + (\sigma_{\rm{exp}}-\sigma^0 _{\rm f}).
\end{equation} 

\item[3)] The steps 1) and 2) are iterated $i$ times giving
\begin{eqnarray}
\sigma^i_{\rm f} &=& {\bf D} \sigma^{i}
\\
\sigma^{i+1}     &=& \sigma^i + (\sigma_{\rm{exp}}-\sigma^i _{\rm f})
\end{eqnarray}
until convergence is achieved. This means that
$\sigma^{i+1}_{\rm f} \approx \sigma_{\rm exp}$ within the statistical errors.
In order to quantitatively check convergence, we calculate the reduced $\chi^2$ of $\sigma^{i+1}_{\rm f}$ and
$\sigma_{\rm{exp}}$ after each iteration.
Approximately four iterations are usually enough for convergence, which is defined when the reduced $\chi^2$ value approaches $\approx 1$.
\end{itemize}

We stopped iterating when the $\chi^2$ started to
be lower than unity. In principle, the iteration could continue until the reduced $\chi^2$ approaches zero,
but that results in large unrealistic fluctuations in $\sigma^i$ due to over-fitting to the measured points $\sigma_{\rm exp}$.
To prevent the unfolding from introducing fluctuations that do not reflect nuclear properties, we apply a smoothing factor of 200-300 keV, which corresponds to the average of the full-width half maximum (FWHM) of the $\gamma$ beams. 

\begin{figure}[]
\includegraphics[clip,width=0.85\columnwidth]{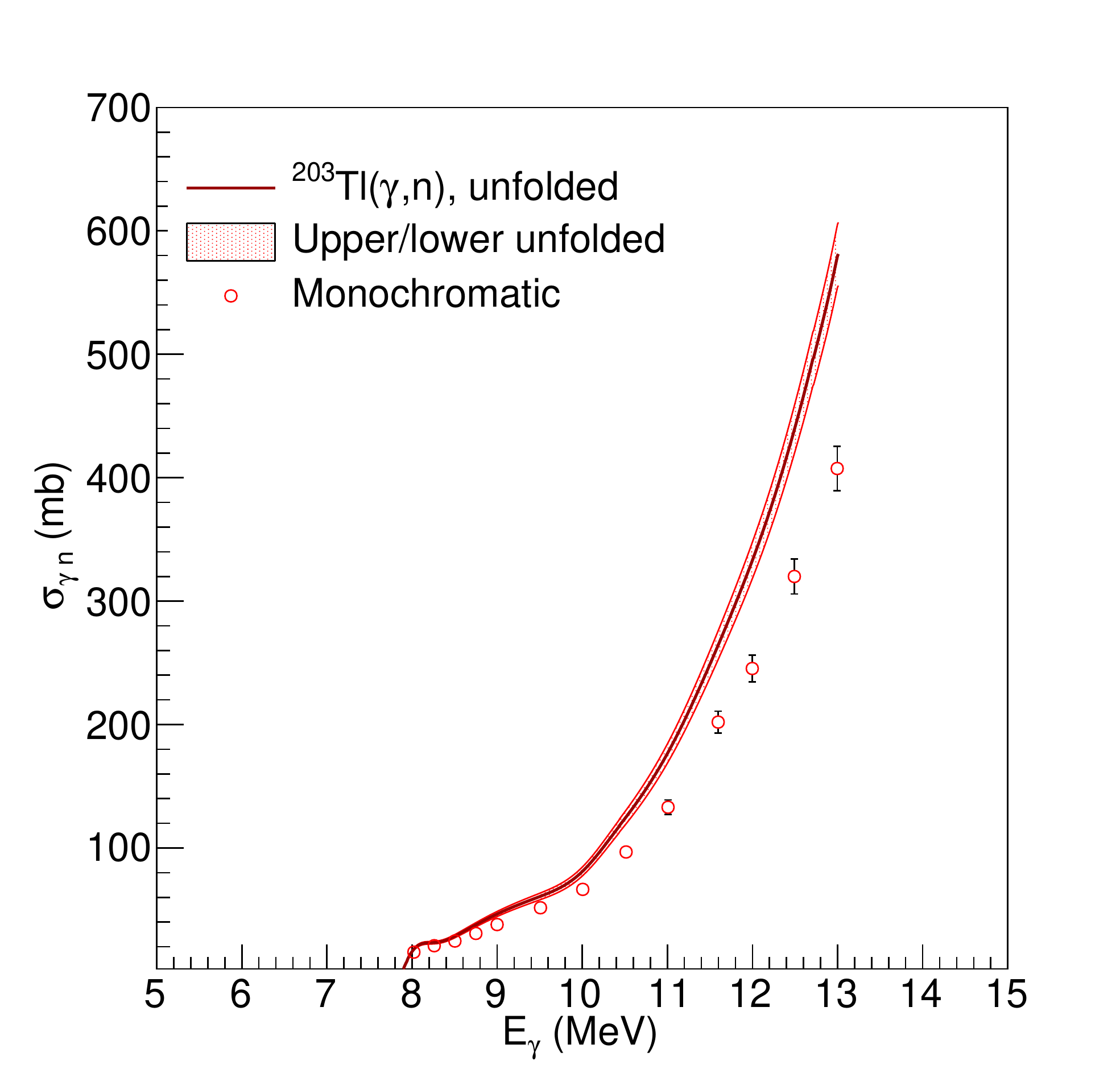}
\caption{(Color online) Cross sections of $^{203}\rm{Tl}$. The green open circles are the monochromatic cross sections from Fig.\ref{fig:MonocrossBoth}. The green, shaded area display the unfolded cross section.}
\label{fig:Unfold_203}
\end{figure}

\begin{figure}[]
\includegraphics[clip,width=0.85\columnwidth]{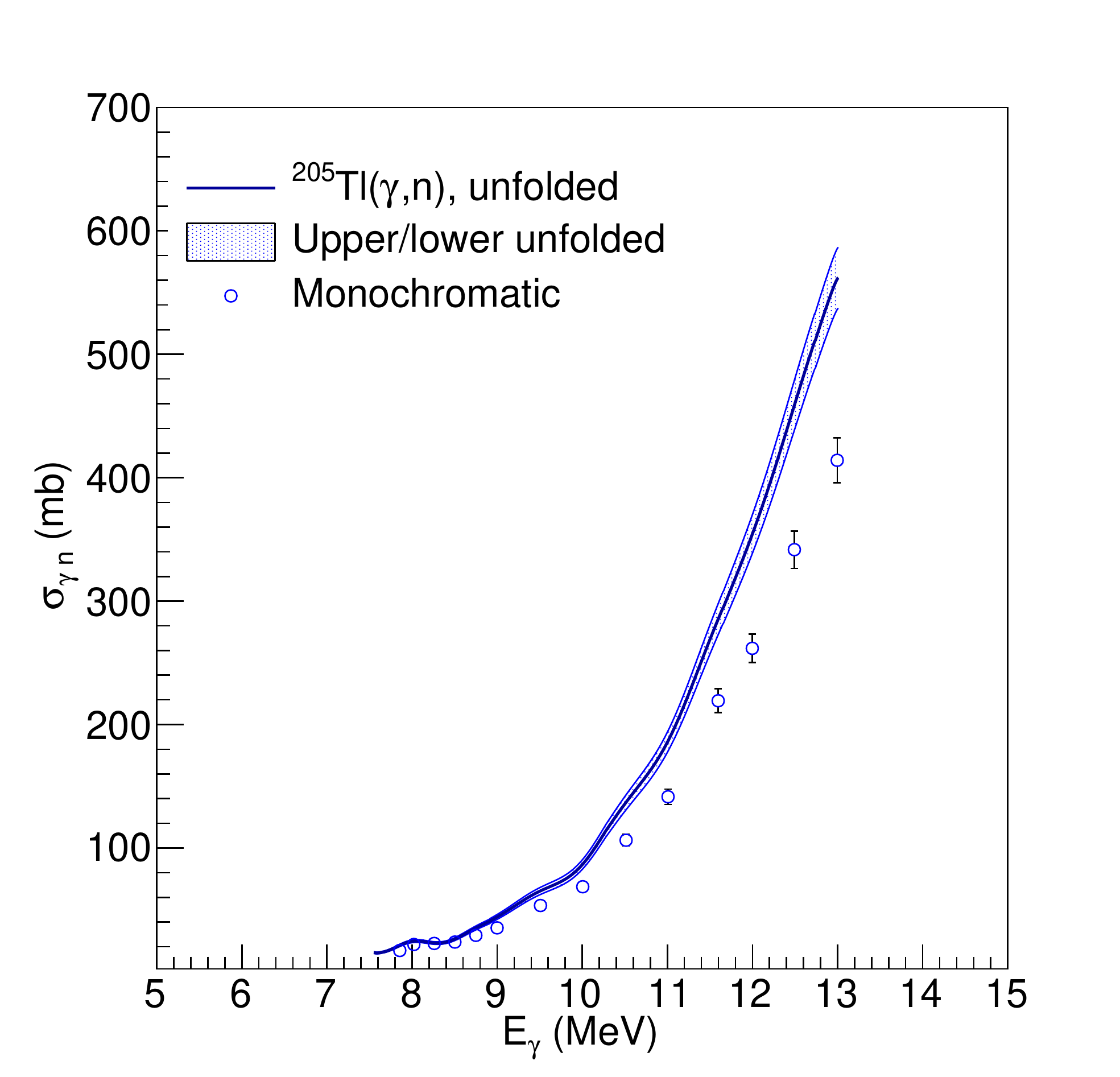}
\caption{(Color online) Cross sections of $^{205}\rm{Tl}$. The blue filled squares are the monochromatic cross sections from Fig.\ref{fig:MonocrossBoth}. The blue, shaded area display the unfolded cross section.}
\label{fig:Unfold_205}
\end{figure}

We estimate the total uncertainty in the unfolded cross sections by calculating an upper limit of the monochromatic cross sections from Fig.\ref{fig:MonocrossBoth} by adding and subtracting the errors to the measured cross section values. This upper and lower limit is then unfolded separately, resulting in the unfolded cross sections shown in Figs.~\ref{fig:Unfold_203} and~\ref{fig:Unfold_205}.

\begin{figure}[]
\includegraphics[clip,width=0.9\columnwidth]{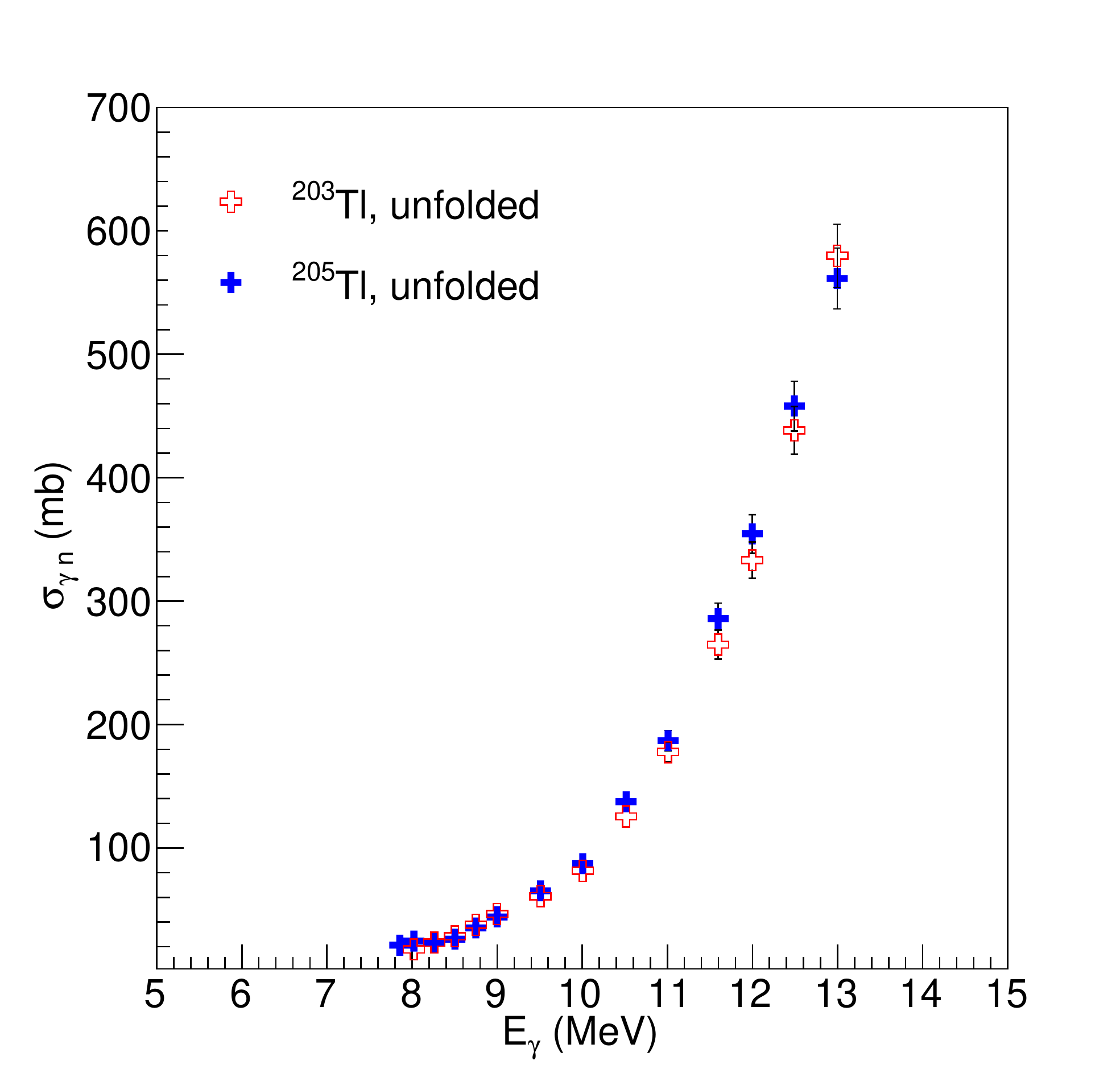}
\caption{(Color online) The recommended unfolded cross sections of $^{203,205}\rm{Tl}$. Here, the error bars represent the difference between the upper and lower limits shown in Figs.~\ref{fig:Unfold_203} and ~\ref{fig:Unfold_205}.}
\label{fig:Unfold_203205}
\end{figure}

In Fig.~\ref{fig:Unfold_203205}, the two unfolded cross sections are evaluated at the maximum energies of the incoming $\gamma$ beams. The error bars represent the difference between the upper and lower limit of the unfolded cross sections.

%---------------------------------------------------%
%---------------------------------------------------%

%---------------------------------------------------%
%---------------------------------------------------%

%---------------------------------------------------%
%---------------------------------------------------%
\section{Discussion}
\label{sec_disc}

%---------------------------------------------------%
\begin{figure}
\includegraphics[width=1.0\columnwidth]{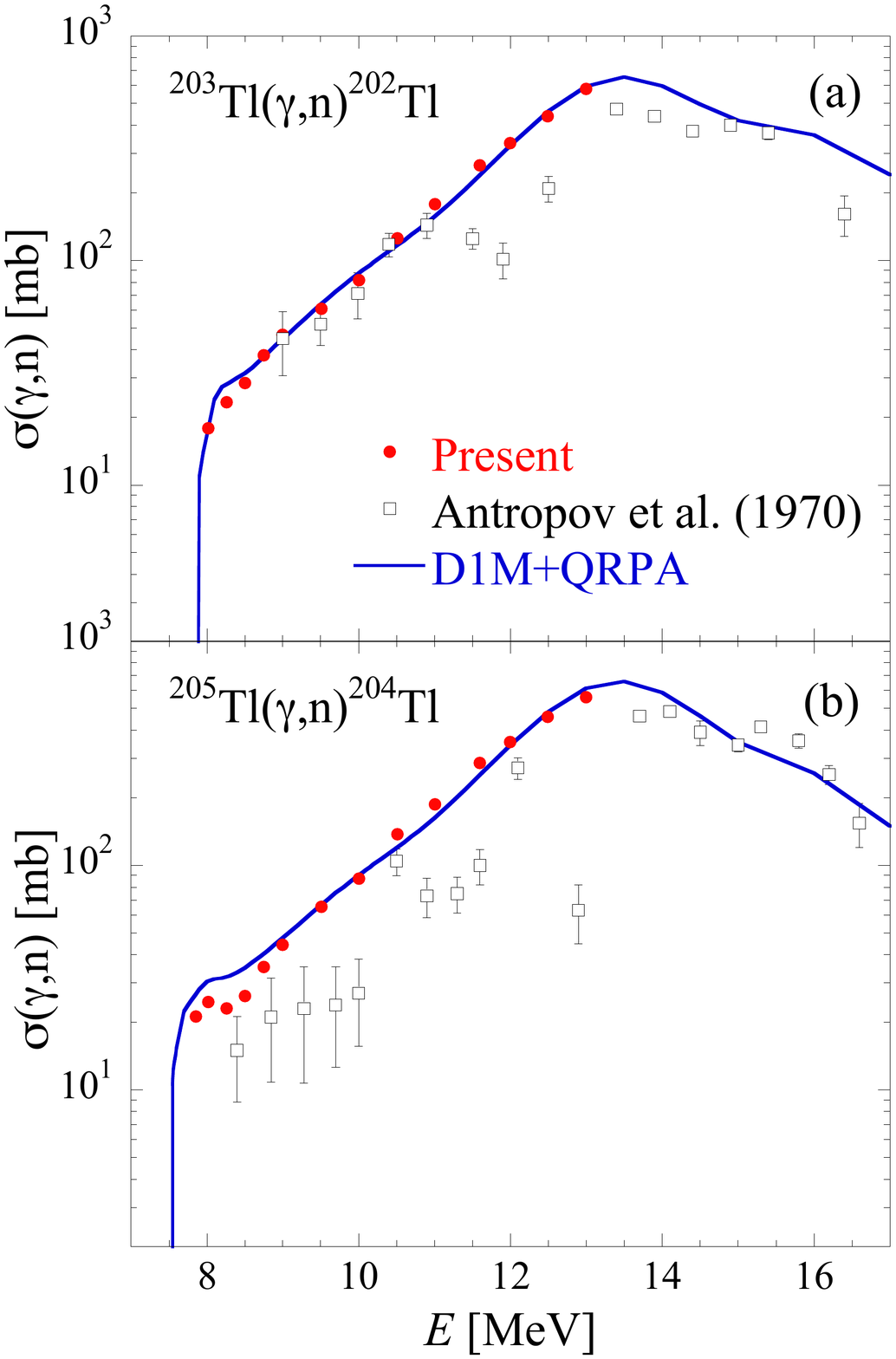}
\caption{(Color online) (a) Present $^{203}\rm{Tl}$($\gamma$,n)$^{202}\rm{Tl}$ measured cross sections compared with the D1M+QRPA calculations and previous measurements (open squares) \cite{Antropov70}. (b) same for  $^{205}\rm{Tl}$($\gamma$,n)$^{204}\rm{Tl}$. }
\label{fig:gn}
\end{figure}
%---------------------------------------------------%
%---------------------------------------------------%
\begin{figure}
\includegraphics[width=1.0\columnwidth]{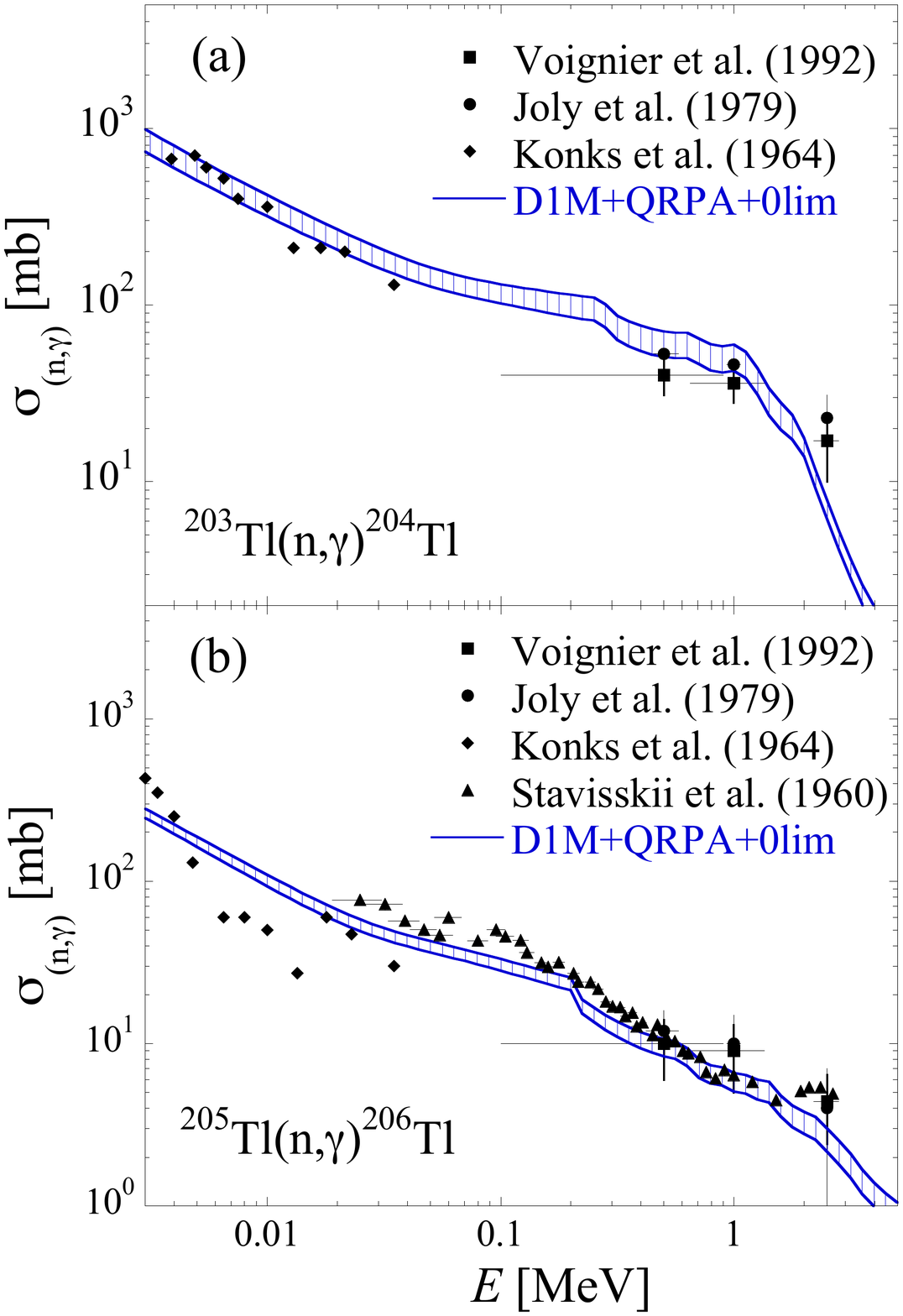}
\caption{(Color online) (a) $^{203}\rm{Tl}$(n,$\gamma$)$^{204}\rm{Tl}$ measured cross section compared with the D1M+QRPA+0lim calculations (blue shaded lines). The hashed area is obtained making use of different nuclear densities. (b) same for  $^{205}\rm{Tl}$(n,$\gamma$)$^{206}\rm{Tl}$. Previous experimental data are taken from Refs.~\cite{Voignier92,Joly79,Konks64,Stavisskii60}. }
\label{fig:ng}
\end{figure}
%---------------------------------------------------%

%---------------------------------------------------%
\begin{figure}
\includegraphics[width=1.0\columnwidth]{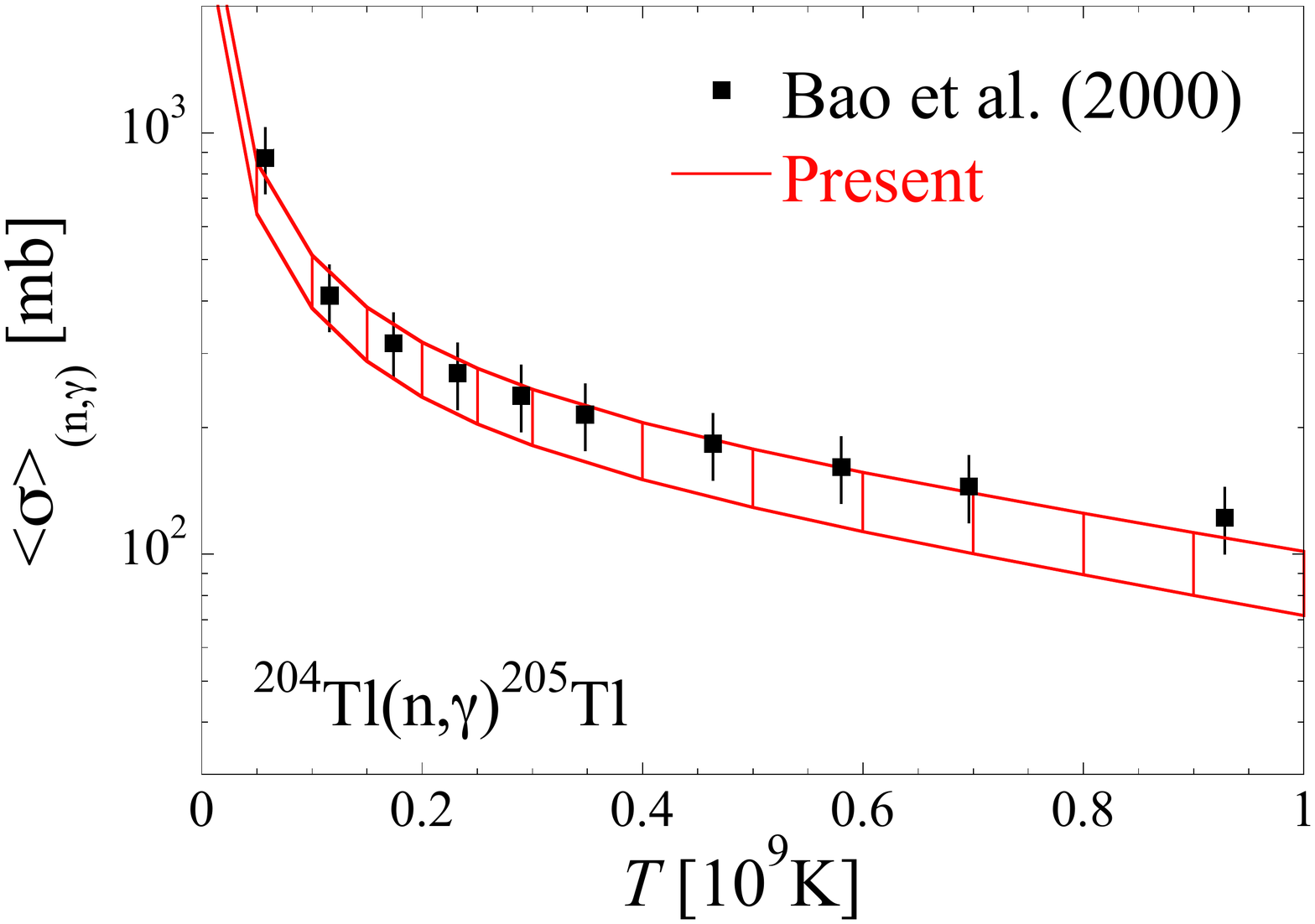}
\caption{(Color online)  $^{204}\rm{Tl}$(n,$\gamma$)$^{205}\rm{Tl}$ Maxwellian-averaged cross section calculated with the D1M+QRPA+0lim strength function (red lines) as a function of the temperature $T$. The hashed area is obtained making use of different nuclear densities. Also included is the recommended cross section of Bao et al. (2000) \cite{Bao00}.}
\label{fig:204Tlng}
\end{figure}
%---------------------------------------------------%

The present experimental results are now analyzed in light of the recent systematics of the $\gamma$SF obtained within the mean field plus QRPA calculations based on the finite-range Gogny D1M interaction \cite{Martini16,Goriely16b,Goriely18a}. When compared with experimental data and considered for practical applications, the mean field plus QRPA calculations need  some phenomenological corrections. These include a broadening of the QRPA strength  to take the neglected damping of  collective motions into account as well as a shift of the strength to lower energies due to the contribution beyond the 1 particle - 1 hole excitations and the interaction between the single-particle and low-lying collective phonon degrees of freedom. Such phenomenological corrections  have been applied to the present Tl isotopes, as described in Ref.~\cite{Goriely18a}. In addition, in order to reproduce the present photoneutron cross section in the low-energy tail of the giant dipole resonance (GDR), we find that a global energy shift of 0.7~MeV of the overall $E1$ strength and a reduction factor of 2 on the $M1$ strength are required. Such renormalizations are within the uncertainties affecting the $\gamma$SF predictions \cite{Goriely18a} and are applied to all the Tl $\gamma$SF studied in the present work.

The resulting D1M+QRPA photoneutron cross section calculated with the TALYS reaction code \cite{Koning12} is shown in Fig.~\ref{fig:gn} for both $^{203}\rm{Tl}$($\gamma$,n)$^{202}\rm{Tl}$ and $^{205}\rm{Tl}$($\gamma$,n)$^{204}\rm{Tl}$. Although the fit is not perfect, D1M+QRPA calculation is seen to reproduce fairly well the dipole strength measured in the present study in the 8--13~MeV region. In contrast, major differences with the previous measurements \cite{Antropov70} can be observed, especially in the 11--13~MeV range, where significantly lower cross sections were extracted from this bremsstrahlung experiment. These former data were used to estimate the GDR peak cross section $\sigma_r$, peak energy $E_r$, and width at half maximum $\Gamma_r$ in photoabsorption studies \cite{Plujko18}. For $^{203}\rm{Tl}$, $\sigma_r=437 \pm94$~mb, $E_r=14.06\pm 0.08$~MeV and $\Gamma_r=3.95\pm 0.21$~MeV were deduced, and for $^{205}\rm{Tl}$, $\sigma_r=479 \pm16$~mb, $E_r=14.47\pm 0.05$~MeV and $\Gamma_r=2.93\pm 0.16$~MeV. Based on our new measurements (Fig.~\ref{fig:gn}), we find GDR parameters corresponding to $\sigma_r=655$~mb, $E_r\simeq 13.5$~MeV, $\Gamma_r\simeq3.2$~MeV for $^{203}\rm{Tl}$ and $\sigma_r=660$~mb, $E_r \simeq 13.5$~MeV and $\Gamma_r\simeq3.5$~MeV for $^{205}\rm{Tl}$. Our lower value of the GDR peak energies lead to a $\gamma$SF at low energies significantly larger in comparison with what was extracted from the Antropov et al. data. 

Another way of testing our photoneutron data is to consider the reverse radiative neutron capture cross sections. Those are also available for $^{203}\rm{Tl}$ and $^{205}\rm{Tl}$, but depend on the de-excitation strength function of the compound nuclei  $^{204}\rm{Tl}$ and $^{206}\rm{Tl}$ for which no experimental data exists. Nevertheless, we have considered the D1M+QRPA $E1$ and $M1$ strengths renormlaized in the same way as described above and applied to the calculation of the (n,$\gamma$) cross section. We compare in Fig.~\ref{fig:ng}  the $^{203}\rm{Tl}$(n,$\gamma$)$^{204}\rm{Tl}$ and $^{205}\rm{Tl}$(n,$\gamma$)$^{206}\rm{Tl}$ measured cross sections with the TALYS Hauser-Feshbach calculation based on the D1M+QRPA+0lim strength functions and different nuclear level density prescriptions. All nuclear level densities \cite{Koning08,Goriely08,Hilaire12} are normalized to the existing s-wave spacing data at the neutron binding energy \cite{Capote09}. Also note that the D1M+QRPA photoabsorption strength needs to be complemented by the zero-limit correction when considering the de-excitation of the compound nucleus formed by the neutron capture. Inspired from shell model studies, this low-energy limit has been approximated in Ref.~\cite{Goriely18a} and when complementing the QRPA calculation, the final $\gamma$SF is referred to as D1M+QRPA+0lim. As shown in Fig.~\ref{fig:ng}, the calculated cross sections are in rather good agreement with experimental data in the keV region. 

Such a comparison also increases our confidence on the relevance of our new measurements and the corresponding theoretical D1M+QRPA+0lim $\gamma$SF adjustment and allows us to estimate the radiative neutron capture cross section of the s-process branching point $^{204}\rm{Tl}$. We show in Fig.~\ref{fig:204Tlng} the Maxwellian-averaged cross section predicted with the D1M+QRPA+0lim strength function (red lines) as a function of the temperature $T$. The hashed area reflects the sensitivity of the predictions with respect to different nuclear density models \cite{Koning08,Goriely08,Hilaire12}. We also compare in Fig.~\ref{fig:204Tlng} our predictions with the one recommended by the compilation of Bao et al. ~\cite{Bao00} widely used for nucleosynthesis applications. Both calculations are seen to be compatible. Based on our experimentally constrained cross section, we can therefore confirm previous nucleosynthesis predictions using the $^{204}\rm{Tl}$(n,$\gamma$)$^{205}\rm{Tl}$ rate of Bao et al. \cite{Bao00}, hence the possible impact of the $^{204}\rm{Tl}$ branching point on the s-process production of $^{205}\rm{Tl}$ and $^{205}\rm{Pb}$.

\section{Conclusion}
\label{sec_conc}
We presented a new experimental determination of the $(\gamma,n)$ cross section for $^{203}$Tl and $^{205}$Tl performed at the NewSUBARU synchrotron radiation facility. Our new measurements cover the low-energy tail of the GDR above the neutron threshold and give significantly different cross sections in comparison with the previous bremsstrahlung experiment of Ref.~\cite{Antropov70}. The GDR parameters have been re-estimated leading to significantly lower GDR peak energies and larger peak cross sections. The new cross sections have been used to constrain the E1 and M1 strength functions obtained within the D1M+QRPA approach. We have further confirmed the relevance of the experimentally constrained D1M+QRPA dipole $\gamma$-ray strength function by analyzing the radiative neutron capture cross sections for Tl isotopes considering in addition the zero-limit systematics for both the de-excitation E1 and M1 strengths. Finally, the present analysis was used to estimate the Maxwellian-averaged $^{204}\rm{Tl}$(n,$\gamma$)$^{205}\rm{Tl}$ cross section, which is of direct relevance to the $^{205}$Pb -- $^{205}$Tl chronometry, which can now be considered to be rather reliably determined.

\section{Acknowledgments}
The authors are grateful to H. Ohgaki of the Institute of Advanced Energy, Kyoto University for making a large volume LaBr$_3$(Ce) detector available for the experiment. H.U. acknowledges the support from the Premier Project of the Konan University. S.G. acknowledges the support from the F.R.S.-FNRS. G.M.T. acknowledges funding from the Research Council of Norway, Project Grant Nos. 262952.  A.C.L. acknowledges funding from ERC-STG-2014, grant agreement No. 637686.  This work was supported by the IAEA and performed within the IAEA CRP on ``Updating the Photonuclear data Library and generating a Reference Database for Photon Strength Functions'' (F41032).

\bibliographystyle{apsrev4-1}
\bibliography{newsubarubibfile}

\begin{thebibliography}{}
\bibitem{Bartholomew73} G.A. Bartholomew, E.D. Earle, A.J. Fergusson, J.W. Knowles, mad M.A. Lone, Adv. Nucl. Phys. {\bf 7}, 229 (1973).
\bibitem{Lone85} M.A. Lone, Proc. 4th Int. Symp., Smolenice, Czechoslovakia, 1985, J. Kristin, E. Betak (eds.), D. Reidel, Dordrecht, Holland (1986) 238.
\bibitem{RIPL3} R. Capote, M. Herman, P. Oblo\u zinsk\'y, P.G. Young, S. Goriely, T. Belgya, A.V. Ignatyuk, A.J. Koning, S. Hilaire, V.A. Plujko, M. Avrigeanu, O. Bersillon, M.B. Chadwick, T. Fukahori, Zhigang Ge, Yinlu Han, S. Kailas, J. Kopecky, V.M. Maslov, G. Reffo, M. Sin, E.Sh. Soukhovistskii, P. Talou, Nuclear Data Sheets {\bf 110}, 3107 (2009). 
\bibitem{Goriely18a} S. Goriely, S. Hilaire, S. P\'eru, K. Sieja, Phys. Rev. C 98 (2018) 014327.
%\bibitem{Gori18} S. Goriely, S. Hilaire, S. P\'{e}ru, and K. Sieja, Phys. Rev. C {\bf 98}, 014327 (2018).
\bibitem{Brink} D.M. Brink, Ph.D thesis, Oxford University, 1955.
\bibitem{Axel} P. Axel, Phys. Rev. {\bf 126}, 671 (1962).
\bibitem{Voin04}A. Voinov, E. Algin, U. Agvaanluvsan, T. Belgya, R. Chankova, M. Guttormsen, G. E. Mitchell, J. Rekstad, A. Schiller, and S. Siem, 
Phys. Rev. Lett. {\bf 93}, 142504 (2004).
\bibitem{Gutt05} M. Guttormsen, R. Chankova, U. Agvaanluvsan, E. Algin, L. A. Bernstein, F. Ingebretsen, T. L\"onnroth, S. Messelt, 
G. E. Mitchell, J. Rekstad, A. Schiller, S. Siem, A. C. Sunde, A. Voinov, and S. {\O}deg{\aa}rd, Phys. Rev. C {\bf 71}, 044307 (2005).
\bibitem{Algi08} E. Algin, U. Agvaanluvsan, M. Guttormsen, A. C. Larsen, G. E. Mitchell, J. Rekstad, A. Schiller, S. Siem, and A. Voinov,
Phys. Rev. C {\bf 78}, 054321 (2008).
\bibitem{Schw13} R. Schwengner, S. Frauendorf, and A. C. Larsen, Phys. Rev. Lett. {\bf 111}, 232504 (2013).
\bibitem{Brow14} B. A. Brown and A. C. Larsen, Phys. Rev. Lett. {\bf 113}, 252502 (2014).
\bibitem{Siej17a} K. Sieja, Phys. Rev. Lett. {\bf 119}, 052502 (2017).
\bibitem{Siej17b} K. Sieja, Europhys. J. Web Conf. {\bf 146}, 05004 (2017).
\bibitem{Kara17} S. Karampagia, B. A. Brown, and V. Zelevinsky, Phys. Rev. C {\bf 95}, 024322 (2017).
\bibitem{Schw17} R. Schwengner, S. Frauendorf, and B. A. Brown, Phys. Rev. Lett. {\bf 118}, 092502 (2017).
\bibitem{Mowl89} N. Mowlavi, S. Goriely, and M. Arnould, Astron. Astrophys. {\bf 330}, 206 (1998).
\bibitem{Arno97} M. Arnould, G. Paulus, G. Meynet, Astron. Astrophys. {\bf 321}, 452 (1997).
\bibitem{Ande60} E. Anders, C.M. Stevens, J. Geophys. Res. {\bf 65}, 3043 (1960).
\bibitem{Osti69} R.G. Ostic, H.M. El-Badry, T.P. Kohman, Earth Planetary Sci. Letters {\bf 7}, 72 (1969).
\bibitem{Huey72} J.M. Huey, T.P. Kohman, Earth Planetary Sci. Letters {\bf 16}, 401 (1972).
\bibitem{Taka83} K. Takahashi, K. Yokoi, Nucl. Phys. A {\bf 404}, 578 (1983).
\bibitem{Daud47} R. Daudel, P. Benoist, R. Jacques, M. Jean, Compt. rend. {\bf 224}, 1427 (1947).
\bibitem{Bahc61} J.N. Bahcall, Phys. Rev. {\bf 124}, 495 (1961).
\bibitem{Yoko85} K. Yokoi, K. Takahashi, M. Arnould, Astron. Astrophys. {\bf 145}, 339 (1985).
\bibitem{Utsu10a} H. Utsunomiya, S. Goriely, H. Akimune, H. Harada, F. Kitatani, S. Goko, H. Toyokawa, K. Yamada, T. Kondo, O. Itoh, M. Kamata, T. Yamagata, 
Y.-W. Lui, I. Daoutidis, D. P. Arteaga, S. Hilaire, and A. J. Koning, Phys. Rev. C {\bf 82}, 064610 (2010). 
\bibitem{Utsu10b} H. Utsunomiya, S. Goriely, H. Akimune, H. Harada, F. Kitatani, S. Goko, H. Toyokawa, K. Yamada, T. Kondo, O. Itoh, M. Kamata, T. Yamagata, 
Y.-W. Lui, S. Hilaire, and A. J. Koning, Phys. Rev. C {\bf 81}, 035801 (2010). 
\bibitem{Utsu11} H. Utsunomiya, S. Goriely, M. Kamata, H. Akimune, T. Kondo, O. Itoh, C. Iwamoto, T. Yamagata, H. Toyokawa, Y.-W. Lui, H. Harada, F. Kitatani, 
S. Goko, S. Hilaire, and A. J. Koning, Phys. Rev. C {\bf 84}, 055805 (2011).
\bibitem{Utsu13} H. Utsunomiya, S. Goriely, T. Kondo, C. Iwamoto, H. Akimune, T. Yamagata, H. Toyokawa, H. Harada, F. Kitatani, Y.-W. Lui, A. C. Larsen, 
M. Guttormsen, P. E. Koehler, S. Hilaire, S. P?ru, M. Martini, and A. J. Koning, Phys. Rev. C {\bf 88}, 015805 (2013).
\bibitem{Fili14} D. M. Filipescu, I. Gheorghe, H. Utsunomiya, S. Goriely, T. Renstr\o m, H.-T. Nyhus, O. Tesileanu, T. Glodariu, T. Shima, K. Takahisa, S. Miyamoto, Y.-W. Lui, S. Hilaire, S. P\'eru, M. Martini, and A. J. Koning, Phys. Rev. C {\bf 90}, 064616 (2014).
\bibitem{Nyhu15} H.-T. Nyhus, T. Renstr{\o}m, H. Utsunomiya, S. Goriely, D. M. Filipescu, I. Gheorghe, O. Tesileanu, T. Glodariu, T. Shima, K. Takahisa, S. Miyamoto, Y.-W. Lui, S. Hilaire, S. P\'eru, M. Martini, L. Siess, and A. J. Koning, Phys. Rev. C {\bf 91}, 015808 (2015). 
%Begin experimental part
\bibitem{Utsu14} H.~Utsunomiya, T. Shima, K. Takahisa, D.M. Filipescu, O. Tesileanu, I. Gheorghe, H.-T. Nyhus, T. Renstr\o m, Y.-W. Lui, Y. Kitagawa, S. Amano, S. Miyamoto, IEEE Trans. Nucl. Sci.  {\bf 61}, 1252 (2014).
\bibitem{Ioana_thesis} A. I. Gheorghe, PhD thesis: Nuclear data obtained with Laser Compton Scattered gamma-ray beams, Ph.D. thesis, University of Bucharest (2017), unpublished.
\bibitem{geant4ref}  J. Allison {\it et al.}, IEEE T. Nucl. Sci. {\bf 53}, 270 (2006). 
\bibitem{neutrondet} O. Itoh, H. Utsunomiya, H. Akimune, T. Kondo, M. Kamata, T. Yamagata, H. Toyokawa, H. Harada, F. Kitatani, S. Goko, C. Nair, and Y.-W. Lui, Journal of Nuclear Science and Technology {\bf 48}, 834 (2011).
\bibitem{Berman_ring_ratio} B. L. Berman, J. T. Caldwell, R. R. Harvey, M. A. Kelly, R. L. Bramblett, and S. C. Fultz, Phys. Rev. 162, 1098 (1967).
\bibitem{Utsu2017} H. Utsunomiya, I. Gheorghe, D. M. Filipescu, T. Glodariu, S. Belyshev, K. Stopani, V. Varlamov, B. Ishkhanov,
S. Katayama, D. Takenaka, T. Ari-izumi, S. Amano, S. Miyamoto, Nuclear Instruments and Methods in Physics Research Section A {\bf 871}, 135 (2017).
\bibitem{Kondo2011}T. Kondo, H. Utsunomiya, H. Akimune, T. Yama- gata, A. Okamoto, H. Harada, F. Kitatani, T. Shima, K. Horikawa, and S. Miyamoto, Nuclear Instruments and Methods in Physics Research Section A {\bf 659}, 462(2011).
\bibitem {utsunomiya2018} H. Utsunomiya, T. Watanabe, T. Ari-izumi, D. Takenaka, T. Araki, K. Tsuji, I. Gheorghe, D. M. Filipescu, S. Belyshev, K. Stopani, D. Symochko, H. Wang, G. Fan, T. Renstr{\o}m, G. M. Tveten, Y.-W. Lui, K. Sugita, S. Miyamoto, Nuclear Instruments and Methods in Physics Research Section A {\bf 896},  103 (2018).
%results part
\bibitem{Antropov70} G. P. Antropov, I. E. Mitrofanov, A. I. Prokofev, V. S. Russkikh Izv. Akad. Nauk SSSR, Ser. Fiz 34 (1970) 116
(Bull. Acad. Sci. USSR, Phys. Ser. 34 (1970) 108).
\bibitem{Voignier92} J. Voignier, S. Joly, G. Grenier, Nuclear Science and Engineering 112  (1992), 87
\bibitem{Joly79} S. Joly, J. Voignier, G. Grenier, D.M. Drake, L. Nilsson, Nuclear Science and Engineering 70 (1979)
\bibitem{Konks64} V.A. Konks, F.L. Shapiro, Zhurnal Eksperimental`noi i Teoret. Fiziki 47 (1964) 795
\bibitem{Stavisskii60} Yu.Ya. Stavisskii, V.A. Tolstikov, Atomnaya Energiya 9 (1960) 401

\bibitem{Martini16} M. Martini, S. P\'eru, S. Hilaire, S. Goriely, and F. Lechaftois, Phys. Rev. C {\bf 94}, 014304 (2016).
\bibitem{Goriely16b} S. Goriely, S. Hilaire, S. P\'eru, M. Martini, I. Deloncle, and F. Lechaftois, Phys. Rev. C {\bf 94}, 044306 (2016).
%\bibitem{Goriely18a} S. Goriely, S. Hilaire, S. P\'eru, K. Sieja, Phys. Rev. C 98 (2018) 014327.
\bibitem{Koning12} A.J. Koning, D. Rochman, Nuclear Data Sheets {\bf 113}, 2841 (2012).
\bibitem{Plujko18} V.A. Plujko, O.M. Gorbachenko, R. Capote, P. Dimitriou, Atomic Data and Nuclear Data Tables 123 (2018) 1

\bibitem{Koning08} A.J. Koning, S. Hilaire, S. Goriely, Nucl. Phys.  A {\bf 810}, 13 (2008).
\bibitem{Goriely08} S. Goriely, S. Hilaire, and A.J. Koning, Phys. Rev. C {\bf 78}, 064307 (2008).
\bibitem{Hilaire12} S. Hilaire, M. Girod, S. Goriely, and A.J. Koning, Phys. Rev. C {\bf 86}, 064317 (2012).
\bibitem{Capote09} R. Capote, M. Herman, P. Oblozinsky, {\it et al.}, Nuclear Data Sheets {\bf 110}, 3107 (2009).
\bibitem{Bao00} Z.Y. Bao, H. Beer, F. K\"appeler, F. Voss, K. Wisshak, T. Rauscher, At. Data Nucl. Data Tables {\bf 75}, 1 (2000).

\end{thebibliography}

\end{document}